\documentclass[3p,times]{elsarticle}

\usepackage{ecrc}

\volume{00}

\firstpage{1}

\journalname{}

\runauth{}

\jid{procs}

\jnltitlelogo{}

\usepackage{amssymb}

\usepackage[figuresright]{rotating}

\usepackage{multicol}
\usepackage{physics}
\usepackage{mathdots}

\newcommand{\CRM}{Centre de recherches mathématiques, Université de Montréal\\
P.O. Box 6128, Centre-ville Station, Montréal (Québec), H3C 3J7, Canada}

\newcommand{\DepPhys}{Département de Physique, Université de Montréal, Montréal (Québec), H3C 3J7, Canada}

\newcommand{\IVADO}{Institut de valorisation des données (IVADO), Montréal (Québec), H2S 3H1, Canada}

\newcommand{\Renmin}{School of Mathematics, Renmin University of China, Beijing, 100872, China}

\begin{document}

\begin{frontmatter}



\dochead{}

\title{A classical model for perfect transfer and fractional revival based on $q$-Racah polynomials}


\author[add:crm,add:depphys]{Hugo Schérer}
\author[add:crm,add:depphys,add:ivado]{Luc Vinet} \label{auth:luc}
\author[add:crm,add:renmin]{Alexei Zhedanov} \label{auth:alexei}

\address[add:crm]{\CRM}
\address[add:depphys]{\DepPhys}
\address[add:ivado]{\IVADO}
\address[add:renmin]{\Renmin}

\begin{abstract}
It is shown how choices based on the $q$-Racah polynomials for the masses and spring constants along a chain give new systems that exactly allow dispersionless end-to-end transmission of a pulse as well as periodic splitting of the initial momentum between the first and last mass. This ``Newton's cradle'' provides a classical analog of quantum spin devices that exhibit perfect state transfer and fractional revival.
\end{abstract}


\end{frontmatter}


\begin{multicols}{2}

\section{Introduction}
\label{sec:intro}

Much effort has been put in the design of quantum spin chains exhibiting perfect state transfer (PST) \cite{BosseVinet_2017, Vinet_HowTo, Kay_2011, Bose_2007}, that is chains allowing the end-to-end transfer of a quantum state. The main interest in such quantum wires has to do with the fact that the state transport is realized by the dynamics of the device thereby minimizing the need for external controls and their decoherence effects.

On theoretical grounds and for implementation purposes, it is most useful to have exact models with PST. A number of such analytic ``blueprints'' have been obtained \cite{Vinet_HowTo, Albanese_2004, Chakrabarti_2010, Christandl_2004, Jafarov_2010, Vinet_2012, paraKrawtchouk} using properties of orthogonal polynomials \cite{koekoek} in an inverse spectral problem context \cite{Gladwell}.

Of significant interest also is the classical version of the problem which is to determine if it is possible to construct a chain made out of properly chosen masses and springs so that a pulse given to the first mass will get entirely transmitted to the last one without dispersion. A positive answer to this question was provided recently in \cite{Vaia_Matrix, Vaia_NewtonCradle} where an analytic mass-spring chain exhibiting that property was characterized with the help of the dual Hahn polynomials.

Besides PST, there is another phenomenon that has been explored in quantum spin chains owing to its connection with entanglement generation. It goes under the name of fractional revival (FR) and corresponds to the periodical replication at specific sites of an initial state. Again, orthogonal polynomials have played an important role in determining quantum spin chains with FR \cite{Lemay_FR_and_paraRacah, Genest_FR, Genest_2016}. These studies have in fact provided natural applications for functions known as para-polynomials \cite{paraKrawtchouk, paraRacah, Lemay_q-para-racah} which were otherwise considered as being quite exotic.

The classical enactment of fractional revival was looked at recently in \cite{SVZ} where are presented mass-spring chains showing periodical dispersionless distribution to the first and last masses only, of the momentum initially given to the first mass. As a matter of fact, we called upon the para-Racah polynomials to obtain such analytic systems that have in some cases both perfect transfer and fractional revival and in other cases only the latter.

We present here another family of dispersinoless ``Newton's cradles'' which have the distinctive feature of being based on the $q$-Racah polynomials that sit at the top of the (discrete part of the) Askey scheme. These mass-spring chains are somehow related to a quantum spin chain designed in \cite{Vinet_HowTo} but prove different in their connection to Askey grids or spectra.

The paper will unfold as follows. The relevant dynamics of a mass-spring chain will be reviewed in section \ref{sec:PST_MSC} and the conditions on the normal mode frequencies for perfect transfer to occur will be identified. A striking difference regarding these spectral conditions between the quantum and classical realms is that in the latter case the eigenvalues of the Jacobi matrix that determines the masses and spring constants must be squares of integers. We shall use the recurrence relation obeyed by the $q$-Racah polynomials to identify Jacobi matrices verifying the appropriate conditions and hence obtain systems with perfect transfer. The restrictions on the parameters of the $q$-Racah polynomials to have a mirror symmetric situation and a spectrum made out of squares will be obtained in section \ref{sec:specialQRacah}. The properties of the resulting mass-spring chains will be analyzed in section \ref{sec:solution}. The occurrence of perfect transfer and fractional revival will be studied and we shall also indicate how FR can be swapped with perfect transfer by modifying chains through isospecttral deformations. The paper will end with a summary and concluding remarks.

\section{Perfect transfer and mass-spring chains}
\label{sec:PST_MSC}

We follow the approach in \cite{Vaia_NewtonCradle} to present the problem. A mass-spring chain is completely characterized by the values of the $N+1$ masses $\{m_i\}_{i=0}^N$ and the $N+2$ spring constants $\{K_i\}_{i=0}^{N+1}$ that connect them. Depending on the boundary conditions, the chain is said to be free-free ($K_0 = K_{N+1} = 0$), fixed-fixed ($K_0 \neq 0 \neq K_{N+1}$) or fixed-free ($K_0 \neq 0$ and $K_{N+1} = 0$, or vice versa). Defining $P_i$ as the momentum of $i$-th mass and $Q_i$ as its displacement from equilibrium, the Hamiltonian of this system is given by:
\begin{equation}
    \mathcal{H} = \sum_{i=0}^N \frac{P_i^2}{2m_i} + \frac{1}{2} \sum_{i=0}^{N+1} K_i \qty(Q_{i-1} - Q_i)^2
\end{equation}

\noindent where we define $Q_{-1} = Q_{N+1} = 0$. It is convenient to represent this Hamiltonian with the help of vectors in $\mathbb{R}^{N+1}$  and matrices in $\mathbb{R}^{(N+1)\cp(N+1)}$. Defining $P$ and $Q$ as vectors with entries $P_i$ and $Q_i$ respectively, $M$ as the diagonal mass matrix with entries $M_{ij} = m_i \delta_{ij}$, and $K$ as the following Jacobi (i.e., tridiagonal symmetric) matrix:
\begin{equation}
    K = \mqty(
    K_0 + K_1 & -K_1 & 0 &  \\
    -K_1 & K_1 + K_2 & -K_2 &  \\
    0 & -K_2 & K_2 + K_3 & \ddots\\
     & & \ddots & \ddots & -K_N\\
    &&& -K_N & K_N + K_{N+1}
    ),
\end{equation}

\noindent the Hamiltonian can be written as
\begin{equation}
    \mathcal{H} = \frac{1}{2} P^T M^{-1} P + \frac{1}{2} Q^T K Q.
\end{equation}

\noindent We further define the mass-weighted coordinates and momenta:
\begin{equation}
    q = M^{1/2}Q, \quad p = M^{-1/2}P.
\end{equation}

\noindent The use of $q$ for the mass-weighed displacement should not lead to confusion with the parameter of the $q$-Racah polynomials as the context will always make the intent clear. The Hamiltonian then reads
\begin{equation}
    \mathcal{H} = \frac{1}{2} p^T p + \frac{1}{2} q^T A q,
\end{equation}

\noindent with $A = M^{-1/2} K M^{-1/2}$, the Jacobi matrix given by
\begin{equation}
    A = \mqty(
    b_0 & -\sqrt{u_1} & 0 & \\
    -\sqrt{u_1} & b_1 & -\sqrt{u_2} &  \\
    0 & -\sqrt{u_2} & b_2 & \ddots\\
     & & \ddots & \ddots & -\sqrt{u_N}\\
    &&& -\sqrt{u_N} & b_N
    )_{N+1},
    \label{eq:mat_A}
\end{equation}

\noindent where
\begin{align}
    b_i &= \frac{K_i + K_{i+1}}{m_i}, \quad i = 0,...,N,
    \label{eq:bi_matrix}\\
    \sqrt{u_i} &= \frac{K_i}{\sqrt{m_{i-1}m_i}}, \quad i = 1,...,N.
    \label{eq:ui_matrix}
\end{align}

\noindent Notice that the system is scale-invariant, since multiplying all the masses and spring constants by a constant will give the same matrix exactly.

To obtain perfect end-to-end transfer, it is necessary for the chain to be mirror-symmetric, i.e., that $m_i = m_{N-i}$ and $K_i = K_{N+1-i}$. An immediate consequence is that the matrix $A$ is persymmetric, i.e., that it is invariant under reflection with respect to the antidiagonal, or equivalently that $b_i = b_{N-i}$ and $u_i = u_{N+1-i}$. Also, this means we will only deal with free-free and fixed-fixed cases. Let $U$ be the orthogonal ($UU^T = I$) matrix that diagonalizes $A$,
\begin{equation}
    UAU^T = D \qq{with} D_{mn} = \delta_{mn} x_n,
\end{equation}

\noindent i.e., the $n$-th line of $U$ is the normalized eigenvector corresponding to eigenvalue $x_n$. With the introduction of the normal-mode coordinates and momenta,
\begin{equation}
    \tilde{q} = U^T q, \quad \tilde{p} = U^T p,
\end{equation}

\noindent the Hamiltonian becomes that of $N+1$ independent oscillators with ``spring constant'' $x_n$
\begin{equation}
    \mathcal{H} = \frac{1}{2} \tilde{p}^T \tilde{p} + \frac{1}{2}\tilde{q}^T D \tilde{q}
    = \frac{1}{2} \sum_{n=0}^N \qty(\tilde{p}_n^2 + x_n \tilde{q}_n^2),
\end{equation}

\noindent with the obvious consequence that
\begin{equation}
    x_n = \omega_n^2,
    \label{eq:xn_wn2}
\end{equation}

\noindent where $\omega_n$ are the normal-mode frequencies of the system. We will assume from now on that the $\omega_n$ are ordered, i.e., $\omega_0 < \omega_1 < \dots < \omega_N$. We can describe explicitly the motion of each mass,
\begin{equation}
    q_i(t) = \sum_{n=0}^N U_{ni} \sum_{j=0}^N U_{nj} \qty[q_j(0) \cos \omega_n t + p_j(0) \frac{\sin \omega_n t}{\omega_n}].
    \label{eq:motion}
\end{equation}

\noindent It is worth pointing out at this point that in the case of a free-free chain, the first eigenfrequency will be $\omega_0 = 0$ to account for the translation mode, with the immediate consequence that the matrix $A$ is singular. Also, in (\ref{eq:motion}), it is then understood that $\frac{\sin \omega_0 t}{\omega_0} \mapsto t$. In the fixed-fixed case, $A$ is invertible and (\ref{eq:motion}) holds exactly as is.

The initial conditions for perfect transfer are
\begin{equation}
    q(0) = (0,0,\dots,0)^T, \quad p(0) = (\bar{p},0,\dots,0)^T.
    \label{eq:init_cond}
\end{equation}

\noindent We are interested in the evolution of the momentum of each mass, given by
\begin{equation}
    p_i(t) = \partial_t q_i(t) = \bar{p} \sum_{n=0}^N U_{ni} U_{n0} \cos \omega_n t.
    \label{eq:pN(t)}
\end{equation}

Perfect transfer is achieved if there exists a time $t^*$ such that
\begin{equation}
    p(t^*) = (0,0,...,0,\pm \bar{p}).
    \label{eq:PST_final}
\end{equation}

\noindent The eigenvectors of $A$ alternate between mirror-symmetric and mirror-antisymmetric ones \cite{Cantoni1976}, i.e.
\begin{equation}
U_{n,N-i} = (-1)^n U_{ni},
\label{eq:alternate}
\end{equation}

\noindent so we have
\begin{equation}
    \frac{p_N(t)}{\bar{p}} = \sum_{n=0}^N U_{n0}^2 \cos(n\pi - \omega_n t).
\end{equation}

\noindent Perfect transfer will be achieved if $p_N(t^*)/\bar{p} = \pm 1$, meaning
\begin{equation}
    n\pi - \omega_nt^* = \text{(even integer)} \cp \pi
\end{equation}

\noindent yielding $p_N(t^*)/\bar{p} = 1$, or
\begin{equation}
    n\pi - \omega_nt^* = \text{(odd integer)} \cp \pi
    \label{eq:oddCond}
\end{equation}

\noindent yielding $p_N(t^*)/\bar{p} = -1$. Equivalently, this amounts to having
\begin{equation}
    \omega_n = \omega k_n,
    \label{eq:wn_wkn}
\end{equation}

\noindent with $\omega = \pi/t^*$ and $k_n$ distinct integers with alternating parity and no common factor. This is equivalent to
\begin{equation}
    \delta_n = k_{n+1} - k_{n}
\end{equation}

\noindent being odd positive integers with no common factor. Another useful way to look at this is to require that

\begin{equation}
    \epsilon_n = k_{n+1} + k_{n-1} = \text{(even integer)}
    \label{eq:epsilon_n}
\end{equation}

\noindent be even (positive) integers, which is equivalent to requiring that $k_{n-1}$ and $k_{n+1}$ have the same parity. As this is in fact a second order recurrence relation, it is also necessary to have as initial conditions that $k_0$ and $k_1$ are integers of different parity.

\section{Special cases of the $q$-Racah polynomials}
\label{sec:specialQRacah}

\subsection{Special persymmetric $q$-Racah polynomials}

A first step in our quest for ``Newton's Cradles'' is to produce a persymmetric tridiagonal matrix. The entries for the Jacobi matrix $A$ diagonalized using the $q$-Racah polynomials $R_n(\mu(x); \alpha, \beta, \gamma, \delta | q)$ are the following, in general
\begin{align}
    b_n &= 1 + \gamma \delta q - (A_n + C_n),
    \label{eq:bn}\\
    u_n &= A_{n-1} C_n,
    \label{eq:un}
\end{align}

\noindent with $A_n$ and $C_n$ defined in \cite{koekoek}. Therefore, persymmetry is achieved if $C_{N-n} = A_n$. Direct computation of $C_{N-n}$ first leads to the condition $(\alpha \beta q)^2 = (q^{-N})^2$. If $\alpha \beta q = q^{-N}$, we introduce a singularity in the denominator of $A_n$ and $C_n$. This can be resolved using limits which give rise to another family of orthogonal polynomials, the $q$-para-Racah polynomials \cite{Lemay_q-para-racah}, which deserve an analysis on their own and are out of the scope of the present paper. The only other option is
\begin{equation}
    \alpha \beta q = - q^{-N}.
    \label{eq:alphaBetaQ}
\end{equation}

\noindent Furthermore, one must choose between three sets of two additional conditions for $C_{N-n}$ to be equal to $A_n$. Together with (\ref{eq:alphaBetaQ}), it means that one of the following groups of restrictions must be picked:
\begin{equation}
    \begin{cases}
    \alpha q = q^{-N} \\
    \beta = -1 \\
    \delta = \gamma
    \end{cases}
    \begin{cases}
    \beta \delta q = q^{-N} \\
    \alpha = - \gamma\\
    \delta = \gamma
    \end{cases}
    \begin{cases}
    \gamma q = q^{-N} \\
    \beta = -\alpha \delta^{-1}\\
    \delta = \alpha^2 q^{N+1}
    \end{cases}.
    \label{eq:paramChoices}
\end{equation}

\noindent These three cases actually correspond each to one of the possible conditions for the Askey-Wilson polynomials to truncate to the finite family of the $q$-Racah polynomials (namely $\alpha q = q^{-N}$, $\beta \delta q = q^{-N}$ or $\gamma q = q^{-N}$) \cite{koekoek}. They correspond to these expressions for $A_n$ and $C_n$
\begin{align}
    A_n &= \frac{(1-\gamma^2 q^{2n+2})(1-q^{2n-2N})}{(1+q^{2n-N})(1+q^{2n-N+1})},
    \label{eq:An}\\
    C_n &= \frac{(1-q^{2n})(\gamma^2 q-q^{2n-2N-1})}{(1+q^{2n-N-1})(1+q^{2n-N})} = A_{N-n},
    \label{eq:Cn}
\end{align}

\noindent if $\alpha q = q^{-N}$ or $\beta \delta q = q^{-N}$, and 
\begin{align}
    A_n &= \frac{(1-\alpha^2 q^{2n+2})(1-q^{2n-2N})}{(1+q^{2n-N})(1+q^{2n-N+1})},\\
    C_n &= \frac{(1-q^{2n})(\alpha^2 q-q^{2n-2N-1})}{(1+q^{2n-N-1})(1+q^{2n-N})} = A_{N-n},
\end{align}

\noindent if $\gamma q = q^{-N}$. Any of these choices is equivalent and we shall continue with the choice of $\alpha q = q^{-N}$ without loss of generality. This yields special $q$-Racah polynomials with only one free parameter that we can call $\tilde{P}_n(\mu(x))$ with
\begin{equation}
    \tilde{P}_n(\mu(x); \gamma | q) = R_n(\mu(x); q^{-N-1}, -1, \gamma, \gamma | q),
\end{equation}

\noindent hence given in terms of the basic hypergeometric function ${}_r \phi_s$ by
\begin{equation}
    \tilde{P}_n(\mu(x); \gamma | q) = {}_4 \phi_3 \qty(\mqty{q^{-n},-q^{n-N},q^{-x},\gamma^2 q^{x+1}\\q^{-N},\gamma q,-\gamma q};q,q).
    \label{eq:P4phi3}
\end{equation}

\noindent The grid $\mu(x)$ is
\begin{equation}
    \mu(x) = q^{-x} + \gamma^2 q^{x+1}.
    \label{eq:mu_x}
\end{equation}

The monic version $P_n(\mu(x)) = A_{n-1} \dots A_0 \tilde{P}_n(\mu(x))$ of these polynomials are orthogonal with respect to 
\begin{equation}
    \sum_{x=0}^N w_x P_m(\mu(x)) P_n(\mu(x)) = u_1 \dots u_n \delta_{mn},
\end{equation}

\noindent with the weights reading
\begin{align}
    w_x = (-q^N)^x 
    \frac{(1-\gamma^2 q^{2x+1})}{(1-\gamma^2 q^{x+1})} 
    &\frac{(q^{-N}, \gamma^2 q^2 ; q)_x}{(q, \gamma^2 q^{N+2} ; q)_x} \nonumber\\
    &\cdot \frac{(\gamma q, -\gamma q ; q)_N}{(-1, \gamma^2 q^2 ; q)_N},
    \label{eq:weights}
\end{align}

\noindent in terms of the $q$-Pochhammer symbol
\begin{equation}
    (a;q)_k = (1-a)(1-aq)...(1-aq^{k-1}),
\end{equation}

\noindent and such that
\begin{equation}
    \sum_{x=0}^N w_x = 1.
\end{equation}

\noindent Finally, the positivity of the $u_n$ requires that
\begin{align}
    \abs{\gamma q} &> q^{-N+1}, \qq{or}
    \label{eq:positivity1}\\
    \abs{\gamma q} &< 1,
    \label{eq:positivity2}
\end{align}

\noindent assuming $0 < q < 1$. Notice how the sign of $\gamma$ is not important, as (\ref{eq:bn}), (\ref{eq:An}), (\ref{eq:Cn}), (\ref{eq:P4phi3}), (\ref{eq:mu_x}) and (\ref{eq:weights}) depend only on $\gamma^2$, or have equivalent terms in $\gamma$ and $-\gamma$. (Note that (\ref{eq:bn}) depends only on $\gamma^2$ because $\delta = \gamma$.)

\subsection{Integer eigenvalues on the hyperbolic lattice}

As explained in \cite{Vinet_HowTo}, it is possible to produce integer eigenvalues with alternating parity on a $q$ hyperbolic analog of the uniform spectrum, for example
\begin{equation}
    \check{x}_n = \check{q}^{-n} + d \check{q}^{n+1}.
\end{equation}

Before going further, let us remember that given a set of eigenvalues $\check{x}_n$, it is possible to perform affine transformations to obtain new eigenvalues $k_n$ and new matrix entries that are still diagonalized by the same matrix $U$,
\begin{equation}
    k_n = \Omega (\check{x}_n + \Delta).
    \label{eq:affineTransfo}
\end{equation}

\noindent This transforms the recurrence coefficients as follows,
\begin{equation}
    b_n = \Omega (\check{b}_n + \Delta), \quad u_n = \Omega^2 \check{u}_n,
    \label{eq:affine_bn_un}
\end{equation}

\noindent and the new monic polynomials $P_n(x)$ with these $b_n$ and $u_n$ as recurrence coefficients are related to the former, $\check{P}_n(x)$, by
\begin{equation}
    P_n(x) = \Omega^n \check{P}_n \qty(\frac{x}{\Omega} - \Delta).
\end{equation}

Also, let us recall that we need integers that are perfect squares. Note then that 
\begin{align}
    (k_n)^2 = \Omega^2 \Big((\check{q}^2)^{-n} + &d^2 (\check{q}^2)^{n+1} + 2 d \check{q} \nonumber \\
    &+ 2 \Delta (\check{q}^{-n} + d \check{q}^{n+1}) + \Delta^2\Big).
\end{align}

\noindent Upon choosing $\Delta = 0$, we recover an hyperbolic grid in $q = \check{q}^2$, with a specific affine transformation involving a shift of $2 d \check{q}$ and a multiplication by $\Omega^2$. We notice that this is actually the grid $\mu(x)$ of the $q$-Racah polynomials, with
\begin{equation}
    d^2 = \gamma \delta.
    \label{eq:dSquared}
\end{equation}

With all of this in mind, we can now construct integers on the $\check{q} = q^{1/2}$ hyperbolic lattice
\begin{equation}
    k_n = \Omega(\check{q}^{-n} + d \check{q}^{n+1}).
\end{equation}

\noindent Following the idea of \cite{Vinet_HowTo}, first observe that
\begin{equation}
    \epsilon_n = k_{n+1} + k_{n-1} = k_n (\check{q}^{-1} + \check{q}).
\end{equation}

\noindent Half of the $k_n$ are going to be odd and in order to respect (\ref{eq:epsilon_n}), we need
\begin{equation}
    \check{q}^{-1} + \check{q} = 2r, \quad r=2,3,4,\dots,
    \label{eq:q_lat_int_cond}
\end{equation}

\noindent where we do not allow $r=1$ to avoid the degenerate case of $\check{q}=1$. Isolating $\check{q}$, we find
\begin{equation}
    \check{q} = r - \sqrt{r^2-1} \quad r = 2,3,4,\dots,
    \label{eq:q(k)}
\end{equation}

\noindent if we require that $0 < \check{q} < 1$; interestingly, we also have an expression for its inverse:
\begin{equation}
    \check{q}^{-1} = r + \sqrt{r^2-1} \quad r = 2,3,4,\dots.
\end{equation}

\noindent We also need the first two eigenvalues of this lattice to be integers of distinct parity,
\begin{align}
    k_0 &= \Omega(1 + d \check{q}) = \qq{(integer)},\\
    k_1 &= \Omega(\check{q}^{-1} + d \check{q}^2) = \qq{(integer)},\\
    k_0 + k_1 &\equiv 1 \mod 2.
\end{align}

\noindent This being a system of two equations with two unknowns, it can be solved for $\Omega$ and $d$ explicitly,
\begin{align}
    \Omega &= \frac{k_1 - k_0 \check{q}}{\check{q}^{-1} - \check{q}},\\
    d \check{q}^2 &= \frac{k_0 - k_1 \check{q}}{k_1 - k_0 \check{q}}.
    \label{eq:d_qCheck}
\end{align}

\noindent Allowing for an additional parameter $\omega$, the eigenvalues of the $q$-Racah polynomials, where $q = \check{q}^2$, are
\begin{equation}
    x_n = \omega_n^2 = (\omega k_n)^2 = \omega^2 \Omega^2 (q^{-n} + d^2 q^{n + 1} + 2d q^{1/2}).
    \label{eq:final_xn}
\end{equation}

\noindent We can require that $k_0$ and $k_1$ have no common factor; if they do, since all eigenintegers can be generated by a homogeneous recurrence relation, they will all have this common factor, and we can factor it in the additional parameter $\omega$. Note that in distinction to the quantum chain example presented in \cite{Vinet_HowTo}, in the classical realm, as seen in (\ref{eq:final_xn}), the grid $x_n$ cannot be a hyperbolic sine.

\section{Characterizations of the Newton's cradles}
\label{sec:solution}

We can now determine the specifications of models with perfect transfer and fractional revival in the class associated to $q$-Racah polynomials.

\subsection{Perfect transfer in the fixed-fixed case}

Combining (\ref{eq:dSquared}) with (\ref{eq:paramChoices}) determines the last parameter as $\gamma = d$, or explicitly, using (\ref{eq:d_qCheck}),
\begin{equation}
    \gamma q = \frac{k_0 - k_1 q^{1/2}}{k_1 - k_0 q^{1/2}},
    \label{eq:gammaQ}
\end{equation}

\noindent which respects the positivity condition (\ref{eq:positivity2}) as long as $k_0 < k_1$. We have these explicit expressions for $q$ and its inverse:
\begin{align}
    q &= (r - \sqrt{r^2-1})^2 \quad r = 2,3,4,\dots,
    \label{eq:Qexplicit}\\
    q^{-1} &= (r + \sqrt{r^2-1})^2
\end{align}

\noindent which also define the parameter $r$. To be clear, one can choose the integers $r$, $k_0$ and $k_1$, from which $q$, $\gamma$ and $\Omega$ are now determined. The parameter $\omega$ remains free. The entries of the matrix are now
\begin{align}
    b_n &= \omega^2 k_0^2 - (A_n + C_n),\\
    u_n &= A_{n-1} C_n,
\end{align}

\noindent where
\begin{align}
    A_n &= \omega^2 \frac{(k_1 - k_0 q^{1/2})^2}{4(r^2-1)} 
    \frac{(1-\gamma^2 q^{2n+2})(1-q^{2n-2N})}{(1+q^{2n-N})(1+q^{2n-N+1})},\\
    C_n &= \omega^2 \frac{(k_1 - k_0 q^{1/2})^2}{4(r^2-1)} \frac{(1-q^{2n})(\gamma^2 q-q^{2n-2N-1})}{(1+q^{2n-N-1})(1+q^{2n-N})}.
\end{align}

\noindent The parameter $\Omega$ has been incorporated in $A_n$ and $C_n$ and replaced by its explicit expression. The eigenvalues are
\begin{equation}
    x_n = \omega_n^2 = \omega^2 \frac{(k_1 - k_0 q^{1/2})^2}{4(r^2-1)} (q^{-n} + \gamma^2 q^{n + 1} + 2\gamma q^{1/2}),
\end{equation}

\noindent and perfect transfer occurs at time $t^* = \pi/\omega$.

In the case of a fixed-fixed system, and for a mirror-symmetric chain, we choose $K_0 = K_{N+1} \neq 0$ as the scaling parameter. From there, the system is completely determined, and we can follow the proof of lemma 2 in \cite{Nylen1997} to construct the matrix $M^{-1/2}$. Because the matrix $A$ is not singular, we can write
\begin{equation}
    A_{ij}^{-1} = \sum_{n=0}^N \frac{1}{x_n} U_{ni} U_{nj}.
\end{equation}

\noindent If we define
\begin{equation}
    \Gamma_i = \sum_{s=0}^{\lfloor \frac{N}{2} \rfloor} \frac{1}{x_{2s}} U_{2s,i} U_{2s,0},
\end{equation}

\noindent normalized expressions for the masses and spring constants are then given by
\begin{align}
    \frac{m_i}{K_0} &= \frac{2}{\Gamma_0} \Gamma_i^2,\\
    \frac{K_i}{K_0} &= \frac{2}{\Gamma_0} \Gamma_{i-1} \Gamma_i \sqrt{u_i}.
\end{align}

\noindent We can rewrite these expressions to have $m_0$ as the scaling parameter,
\begin{align}
    \frac{m_i}{m_0} &= \qty(\frac{\Gamma_i}{\Gamma_0})^2 \label{eq:mi_fixed},\\
    \frac{K_i}{\omega^2 m_0} &= \qty(\frac{\Gamma_{i-1}}{\Gamma_0}) \qty(\frac{\Gamma_i}{\Gamma_0}) \sqrt{\frac{u_i}{\omega^4} } \label{eq:Ki_fixed},\\
    \frac{K_0}{m_0} &= \frac{K_N}{m_0} = \frac{1}{2 \Gamma_0}. \label{eq:K0_fixed}
\end{align}

\noindent Notice that, because of (\ref{eq:alternate}), $\Gamma_{N-i} = \Gamma_i$ and the mirror-symmetry of the chain is confirmed. One can write the diagonalizing matrix in terms of the orthogonal polynomials as
\begin{equation}
    U_{ni} = \frac{\sqrt{w_n} P_i (x_n)}{\sqrt{u_1 ... u_i}}.
    \label{eq:Uni}
\end{equation}

\noindent Using this and owing to the fact that $P_0 (\mu(x)) = 1$, we can express $\Gamma_i$ as
\begin{equation}
    \Gamma_i = \sum_{s=0}^{\lfloor \frac{N}{2} \rfloor} \frac{w_{2s}}{x_{2s}} \frac{P_i(x_{2s})}{\sqrt{u_1 \dots u_i}}. \label{eq:gamma_i}
\end{equation}

\subsection{Fractional revival in the fixed-fixed case}

Fractional revival (FR) refers to dynamics such that periodically the initial momentum is distributed to a limited number of masses on the chain, all other masses having zero momentum. We now find for which of the $q$-Racah chains is FR occuring. First of all, it can be proved using (\ref{eq:alternate}) that
\begin{align}
    \sum_{s=0}^{\lfloor \frac{N}{2} \rfloor} U_{2s,i} U_{2s,k} = \frac{1}{2} (\delta_{ik} + \delta_{i,N-k}),
    \label{eq:sumU2s}\\
    \sum_{s=0}^{\lfloor \frac{N-1}{2} \rfloor} U_{2s+1,i} U_{2s+1,k} = \frac{1}{2} (\delta_{ik} - \delta_{i,N-k}).
    \label{eq:sumU2s1}
\end{align}

\noindent Since the momentum is given by (\ref{eq:pN(t)}), this means that if there exists certain times $\tau$ such that $\cos \omega_n \tau$ does not depend on $n$ explicitly, but only on its parity, then we will have an expression for $p(\tau)$ in terms of $\delta_{i0}$ and $\delta_{iN}$, which will entail fractional revival.

Let $\tau_{\ell, Z} = \frac{\ell}{Z} t^*, \ell = 0,1,2,\dots Z$ where $Z$ is a positive integer. If one finds all the $Z$ such that
\begin{align}
    k_{2s} &\equiv \pm k_0 \mod (2Z), \qq{and}
    \label{eq:k2sModZ}\\
    k_{2s+1} &\equiv \pm k_1 \mod (2Z),
    \label{eq:k2s1ModZ}
\end{align}

\noindent for any number $2s, 2s+1 \in \{0,\dots,N\}$, then one has
\begin{align}
    \cos(\omega_{2s} \tau_{\ell, Z}) &= \cos(\frac{\ell k_0 \pi}{Z}),\\
    \cos(\omega_{2s+1} \tau_{\ell, Z}) &= \cos(\frac{\ell k_1 \pi}{Z}),
\end{align}

\noindent with no dependence on $n$ (or $s$) anymore. Furthermore, the construction of $q$ relies on the fact that
\begin{equation}
    k_n \qty(q^{-1/2} + q^{1/2}) = k_{n+1} + k_{n-1},
\end{equation}

\noindent which can be rewritten here as the recurrence relation
\begin{equation}
    k_{n+1} = (2r) k_n - k_{n-1} \label{eq:k_n_rec}.
\end{equation}

\noindent Using this last equation, and remembering that $k_0$ and $k_1$ are co-prime, it can be shown that $Z$ will be solution of (\ref{eq:k2sModZ}) and (\ref{eq:k2s1ModZ}) if and only if it solves one of these four sets of conditions:
\begin{align}
    &\begin{cases}
        r &\equiv 0 \mod Z, \label{eq:firstZ}
    \end{cases}\\
    &\begin{cases}
        rk_0 &\equiv k_1 \mod Z,\\
        rk_1 &\equiv k_0 \mod Z,
    \end{cases}\\
    &\begin{cases}
        rk_0 &\equiv 0 \mod Z,\\
        rk_1 &\equiv k_0 \equiv -k_0 \mod Z,
    \end{cases}\\
    &\begin{cases}
        rk_0 &\equiv k_1 \equiv -k_1 \mod Z, \\
        rk_1 &\equiv 0 \mod Z.
    \end{cases}
\end{align}

\noindent Thus, finding all $Z$ respecting these conditions will provide all times $\tau_{\ell, Z}$ when fractional revival occurs. The momentum at such times is
\begin{align}
    \frac{p_i(\tau_{\ell, Z})}{\bar{p}} &= \delta_{i0} \cos((k_1 + k_0)\frac{\ell \pi}{2Z})\cos((k_1 - k_0)\frac{\ell \pi}{2Z}) \nonumber \\
    +& \delta_{iN} \sin((k_1 + k_0)\frac{\ell \pi}{2Z})\sin((k_1 - k_0)\frac{\ell \pi}{2Z}).
    \label{eq:FR}
\end{align}

\noindent At $\tau_{Z,Z} = t^*$, it is easily checked that (\ref{eq:FR}) reduces to $p_i(t^*)/\bar{p} = (-1)^{k_0} \delta_{iN}$, which is indeed perfect transfer. The time $\tau_{0,Z} = 0$ is also another special case as it represents the initial condition. Therefore, fractional revival actually happens only at $\tau_{\ell,Z}$ when $\ell = 1, \dots, Z-1$. Interestingly, systems with perfect transfer will always exhibit fractional revival. This is so because for instance there is fractional revival at time $\tau_{1,r}$. Indeed, notice that $Z=r$ will always be a solution of (\ref{eq:firstZ}). Furthermore, $r$ cannot be equal to 1 according to (\ref{eq:Qexplicit}) implying that $\tau_{1,r} \neq t^*$. This observation that FR will always occur contrasts with the other classical models analysed so far \cite{Vaia_NewtonCradle, SVZ}, which could give rise to systems with perfect transfer only, without FR.

\subsection{The free-free case}

The free-free mass-spring chain can be treated for the most as a special case of the fixed-fixed situation. To account for the translation mode, we take $k_0 = 0$, and thus from (\ref{eq:gammaQ}), we get

\begin{equation}
    \gamma q = -q^{1/2},
\end{equation}

\noindent which respects positivity condition (\ref{eq:positivity2}). The eigenintegers can be generated with the homogeneous recurrence relation (\ref{eq:k_n_rec}), so $k_1$ will inevitably be a common factor of all these integers. This forces to choose $k_1 = 1$. The eigenvalues are
\begin{equation}
    x_n = \omega^2 \frac{(q^{-n} + q^{n} -2)}{4(r^2-1)}.
\end{equation}

\noindent The entries of the matrix $A$ are now
\begin{align}
    b_n &= - (A_n + C_n),
    \label{eq:bnFree}\\
    u_n &= A_{n-1} C_n,
    \label{eq:unFree}
\end{align}

\noindent with
\begin{align}
    A_n &= \frac{\omega^2}{4(r^2-1)} 
    \frac{(1-q^{2n+1})(1-q^{2n-2N})}{(1+q^{2n-N})(1+q^{2n-N+1})},\\
    C_n &= \frac{\omega^2}{4(r^2-1)} \frac{(1-q^{2n})(1-q^{2n-2N-1})}{(1+q^{2n-N-1})(1+q^{2n-N})}.
\end{align}

\noindent The weights reduce to
\begin{align}
    w_x = (-q^N)^x (1+q^x)
    &\frac{(q^{-N} ; q)_x}{(q^{N+1} ; q)_x} \nonumber\\
    &\cdot \frac{(q^{1/2}, -q^{1/2} ; q)_N}{(-1, q ; q)_N}.
    \label{eq:weightsFree}
\end{align}

Since one of the eigenvalues is zero, the matrix $A$ is no longer invertible, and a new approach is needed to solve for the masses and spring constants, which, interestingly, yields closed-form expressions in this case. First, we define $y_i$,
\begin{equation}
    y_i = \sqrt{\frac{m_{i+1}}{m_i} u_{i+1}} =
    \frac{K_{i+1}}{m_i}.
    \label{eq:yi}
\end{equation}

\noindent From (\ref{eq:bi_matrix}) and (\ref{eq:ui_matrix}), we derive the following recurrence relation for $y_i$,
\begin{align}
    y_i &= b_i - \frac{u_i}{y_{i-1}},\\
    y_0 &= b_0.
\end{align}

\noindent From (\ref{eq:bnFree}) and (\ref{eq:unFree}), and the fact that $C_0 = 0$, it is easy to see that $-A_i$ satisfies the same recurrence relation as $y_i$ with the same initial condition, i.e.
\begin{equation}
    y_i = -A_i.
\end{equation}

\noindent From (\ref{eq:yi}) and $m_0$ as the scaling parameter, the solution is
\begin{align}
    m_i &= \frac{(A_{i-1} A_{i-2} \dots A_0)^2}{u_i u_{i-1} \dots u_1} m_0, \label{eq:mi_Aiui}\\
    K_i &= -A_{i-1} m_{i-1} \label{eq:Ki_Aiui}.
\end{align}

\noindent In closed form, one gets
\begin{equation}
    \frac{m_i}{m_0} = \frac{(q^{1/2}, -q^{1/2}, q^{-N}, -q^{-N};q)_i}{(q,-q,q^{-N+1/2}, -q^{-N+1/2};q)_i} \cdot
    \frac{1+q^{2i-N}}{1+q^{-N}},
\end{equation}
\begin{equation}
    \frac{K_i}{\omega^2 m_0} = \frac{1}{4(r^2-1)} \frac{(1-q^{2i})(q^{2i-2N-1}-1)}{(1+q^{2i-N-1})(1+q^{2i-N})} \cdot \frac{m_i}{m_0} .
\end{equation}

\noindent Using these analytic expression to construct free-free mass-spring chains will result in systems exhibiting perfect transfer, as long as $q$ respects (\ref{eq:Qexplicit}). 

The analysis of fractional revival remains valid. The cosines reduce to
\begin{align}
    \cos(\omega_{2s} \tau_{\ell, Z}) &= 1,\\
    \cos(\omega_{2s+1} \tau_{\ell, Z}) &= \cos(\frac{\ell \pi}{Z}),
\end{align}

\noindent and the conditions for $Z$ reduce to
\begin{equation}
    r \equiv 0 \mod Z
\end{equation}

\noindent Again, finding all $Z$ respecting this condition will provide all times $\tau_{\ell, Z}$ when fractional revival occurs. The momentum at such times is
\begin{equation}
    \frac{p_i(\tau_{\ell, Z})}{\bar{p}} = \delta_{i0} \cos^2\qty(\frac{\ell \pi}{2Z}) + \delta_{iN} \sin^2\qty(\frac{\ell \pi}{2Z}).
    \label{eq:FR_free}
\end{equation}

\subsection{Isospectral deformation}

We can also obtain chains with fractional revival from those with perfect transfer with the help of isospectral deformations \cite{Genest_persymmetric} in both the free-free or fixed-fixed cases. 

Let us consider the matrix $V$ of size $N+1$:
\begin{equation}
V = \mqty(
    \sin \theta &&&&& \cos \theta \\
    & \ddots &&& \iddots &  \\
    && \sin \theta & \cos \theta && \\
    && \cos \theta & - \sin \theta && \\
    & \iddots &&& \ddots & \\
    \cos \theta &&&&& - \sin \theta
    )
\end{equation}

\noindent for $N$ odd, and
\begin{equation}
V = \mqty(
    \sin \theta &&&&&& \cos \theta \\
    & \ddots &&&& \iddots &  \\
    && \sin \theta & 0 & \cos \theta && \\
    && 0 & 1 & 0 && \\
    && \cos \theta & 0 & - \sin \theta && \\
    & \iddots &&&& \ddots & \\
    \cos \theta &&&&&& - \sin \theta
    )
\end{equation}

\noindent for $N$ even. We see that $V = V^T$ and that $V^2 = I$.

Let $\tilde{A} = VAV$. From here on, symbols with tilde will be associated with the system described by this new matrix $\tilde{A}$, and the symbols without a tilde will be the expressions derived from the system related to the matrix $A$ in the previous sections. Also, let $j$ be the integer such that
\begin{align}
    N &= 2j + 1 \qq{if $N$ is odd,}\\
    N &= 2j \qq{if $N$ is even.}
\end{align}

\noindent Clearly the matrix $\tilde{A}$ will have the same spectrum as $A$. Furthermore, only a few entries in the matrix change. In fact, $\tilde{b}_i = b_i$ and $\tilde{u}_i = u_i$, for all $i$ except
\begin{align}
    \tilde{u}_{j+1} &= u_{j+1} \cos^2(2\theta),\\
    \tilde{b}_j &= b_j + \sqrt{u_{j+1}} \sin(2\theta),\\
    \tilde{b}_{j+1} &= b_j - \sqrt{u_{j+1}} \sin(2\theta),
\end{align}

\noindent for $N$ odd, and
\begin{align}
    \tilde{u}_j = u_j (\cos \theta + \sin \theta)^2,\\
    \tilde{u}_{j+1} = u_j (\cos \theta - \sin \theta)^2,
\end{align}

\noindent for $N$ even.

We can define a new parameter $\alpha$ (not to be confused with the $\alpha$ in the original $q$-Racah polynomials) such that
\begin{align}
    \sin(2\theta) &= 1 - 2 \alpha,\\
    \cos(2\theta) &= 2 \sqrt{\alpha(1 - \alpha)},
\end{align}

\noindent with $0 \leq \alpha \leq 1$. Equivalently, we can write
\begin{align}
    \sin \theta &= \frac{\sqrt{1-\alpha} - \sqrt{\alpha}}{\sqrt{2}},\\
    \cos \theta &= \frac{\sqrt{1-\alpha} + \sqrt{\alpha}}{\sqrt{2}}.
\end{align}

\noindent Notice how choosing $\alpha = \frac{1}{2}$ will lead to $V=R$, with $R$ the matrix with ones on the antidiagonal and zeroes everywhere else ($R_{ik} = \delta_{i,N-k}$). The transformation $RAR$ performs a reflection of $A$ with respect to the antidiagonal, and so in this case $\tilde{A} = A$, because $A$ is persymmetric. In terms of $\alpha$, the new entries can be written as
\begin{align}
    \tilde{u}_{j+1} &= 4 \alpha (1-\alpha) u_{j+1},
    \label{eq:tilde_uj1Odd}\\
    \tilde{b}_j &= b_j + (1-2\alpha) \sqrt{u_{j+1}},
    \label{eq:tilde_bjOdd}\\
    \tilde{b}_{j+1} &= b_j - (1-2\alpha) \sqrt{u_{j+1}},
    \label{eq:tilde_bj1Odd}
\end{align}

\noindent for $N$ odd, and
\begin{align}
    \tilde{u}_j = 2(1-\alpha) u_j,
    \label{eq:tilde_ujEven}\\
    \tilde{u}_{j+1} = 2\alpha u_j,
    \label{eq:tilde_uj1Even}
\end{align}

\noindent for $N$ even. One can show that this is equivalent to having $\tilde{A}_i = A_i$ and $\tilde{C}_{i} = C_i$ for all $i$ except
\begin{align}
    \tilde{A}_j &= 2 \alpha A_j,
    \label{eq:Aj_alpha}\\
    \tilde{C}_{N-j} &= 2(1-\alpha) C_{N-j},
\end{align}

\noindent for both $N$ odd or even, with (\ref{eq:bn}) and (\ref{eq:un}) still holding but with tildes everywhere. Now $\tilde{A}$ is diagonalized by $\tilde{U} = UV$. Indeed, 
\begin{equation}
    UV \tilde{A} V^T U^T = U A U^T = D.
\end{equation}

\noindent Note that if $A$ is diagonalized by $U$, it is also diagonalized by a matrix with entries $(-1)^n U_{ni}$. We shall use this, along with (\ref{eq:alternate}), to find the expressions for $\tilde{U}$ that will be consistent with $U$ when $\alpha = \frac{1}{2}$. Consequently, the new diagonalizing matrix has entries
\begin{equation}
    \tilde{U}_{ni} = \begin{cases}
    U_{ni} \cos \theta + U_{n,N-i} \sin \theta \qq{if} i \leq j\\
    U_{ni} \cos \theta - U_{n,N-i} \sin \theta \qq{if} i > j
    \end{cases}
\end{equation}

\noindent for $N$ odd, and 
\begin{equation}
    \tilde{U}_{ni} = \begin{cases}
    U_{ni} \cos \theta + U_{n,N-i} \sin \theta \qq{if} i \leq j-1\\
    U_{ni} \qq{if} i = j\\
    U_{ni} \cos \theta - U_{n,N-i} \sin \theta \qq{if} i > j
    \end{cases}
\end{equation}

\noindent for $N$ even. In terms of $\alpha$, and using (\ref{eq:alternate}), this yields
\begin{equation}
    \tilde{U}_{ni} = U_{ni} \cdot \begin{cases}
    \sqrt{1+(-1)^n(1-2\alpha)} \hfill \qq{if} i \leq j\\
    \sqrt{1-(-1)^n(1-2\alpha)} \hfill \qq{if} i > j
    \end{cases}
\end{equation}

\noindent for $N$ odd, and
\begin{equation}
    \tilde{U}_{ni} = U_{ni} \cdot \begin{cases}
    \sqrt{1+(-1)^n(1-2\alpha)} \hfill \qq{if} i < j\\
    1 \hfill \qq{if} i=j\\
    \sqrt{1-(-1)^n(1-2\alpha)} \hfill \qq{if} i > j
    \end{cases}
\end{equation}

\noindent for $N$ even. The system is completely determined once we choose $\tilde{K}_0$ and $\tilde{K}_{N+1}$ \cite{Nylen1997}. Choosing $\tilde{K}_0 = K_0$ and $\tilde{K}_{N+1} = \frac{\alpha}{1 - \alpha} \tilde{K}_0$ is the only option that will lead to momentum conservation at times $\tau_{\ell,Z}$ for a fixed-fixed system, since it is not necessarily conserved when the system is fixed. Using this and (\ref{eq:gamma_i}), we get
\begin{equation}
    \tilde{m}_i = \begin{cases}
    m_i \qq{if} i \leq j\\
    \frac{\alpha}{1-\alpha} m_i \qq{if} i > j
    \end{cases}
    \label{eq:mi_alpha_odd}
\end{equation}
\begin{equation}
    \tilde{K}_i = \begin{cases}
    K_i \qq{if} i \leq j\\
    2 \alpha K_i \qq{if} i = j+1\\
    \frac{\alpha}{1-\alpha} K_i \qq{if} i > j+1
    \end{cases}
    \label{eq:Ki_alpha_odd}
\end{equation}

\noindent for $N$ odd, and
\begin{equation}
    \tilde{m}_i = \begin{cases}
    m_i \qq{if} i \leq j-1\\
    \frac{1}{2(1-\alpha)} m_i \qq{if} i = j\\
    \frac{\alpha}{1-\alpha} m_i \qq{if} i > j
    \end{cases}
    \label{eq:mi_alpha_even}
\end{equation}
\begin{equation}
    \tilde{K}_i = \begin{cases}
    K_i \qq{if} i \leq j\\
    \frac{\alpha}{1-\alpha} K_i \qq{if} i > j
    \end{cases}
    \label{eq:Ki_alpha_even}
\end{equation}

\noindent for $N$ even. One can show from (\ref{eq:mi_Aiui}) and (\ref{eq:Ki_Aiui}) that these expressions also hold in the free-free case.

Fractional revival is now seen to happen at the times $\tau_{\ell, Z}$ previously defined. Indeed, we have
\begin{eqnarray}
    \frac{p_i(\tau_{\ell,Z})}{\bar{p}} &= \delta_{i0} \qty[ (1-\alpha)\cos(\frac{\ell k_0 \pi}{Z}) + \alpha \cos(\frac{\ell k_1 \pi}{Z})] \nonumber\\
    & + \delta_{iN} \sqrt{\alpha (1-\alpha)}\qty[ \cos(\frac{\ell k_0 \pi}{Z}) - \cos(\frac{\ell k_1 \pi}{Z})],
\end{eqnarray}

\noindent and at the time at time $\tau_{Z,Z} = t^*$, the momentum is
\begin{equation}
    \frac{p_i(t^*)}{\bar{p}} = (-1)^{k_0} \qty[\delta_{i0}(1-2\alpha) + \delta_{iN}(2 \sqrt{\alpha(1-\alpha)})]
\end{equation}

\noindent It is obvious from that expression that perfect transfer is only possible if $\alpha = \frac{1}{2}$. For a general $\alpha$, the use of isospectral deformation transforms a system with perfect transfer (and fractional revival) into one exhibiting fractional revival only.

\subsection{Spectral surgery}

A procedure called spectral surgery given in \cite{Vinet_HowTo} explains how to generate a new matrix $\check{A}$ that will yield a system with perfect transfer from an original matrix $A$ that has this property. The new matrix $\check{A}$ will be of size $N$ instead of $N+1$, and will have the same spectrum except for the fact that one spectral point, $x_k$, will be removed, hence the name of this procedure. It is based on the Christoffel transform between the sets of weights $w_s$ and $\check{w}_n$:
\begin{equation}
    \check{w}_s = C (x_s - x_k)w_s, \quad n = 0,1,\dots,k-1,k+1,\dots,N,
    \label{eq:weightSurg}
\end{equation}

\noindent where $C$ is a normalization constant. The polynomials $\check{P}_n(x)$ that are orthogonal relative to the weights $\check{w}_n$ are obtained from the original ones $P_n(x)$ by
\begin{equation}
    \check{P}_n(x) = \frac{P_{n+1}(x) - E_n P_{n}(x)}{x - x_k},
\end{equation}

\noindent with
\begin{equation}
    E_n = \frac{P_{n+1}(x_k)}{P_{n}(x_k)},
\end{equation}

\noindent and the entries of $\check{A}$ are related to those of $A$ by
\begin{align}
    \check{u_n} &= u_n \frac{E_n}{E_{n-1}},\\
    \check{b_n} &= b_{n+1} + E_{n+1} - E_n.
\end{align}

However, the positivity of the weights is preserved only if $k=0$ or $k=N$. To remove other eigenvalues, the trick is actually to remove a neighbouring pair of spectral points with
\begin{equation}
    \check{w}_s = C (x_s - x_k)(x_s - x_{k+1})w_s.
    \label{eq:weightSurg2}
\end{equation}

\noindent The removal of a consecutive pair of spectral points preserves the alternating of parity in the new eigenvalues, as well as the persymmetry in $\check{A}$ \cite{Vinet_HowTo}. The polynomials that are orthogonal with respect to (\ref{eq:weightSurg2}) are obtained by applying the corresponding Christoffel transform twice. In fact, one can apply iteratively this removal of pairs of spectral points, accompanied by the appropriate Christoffel transforms, an arbitrary number of times to construct a new matrix of desired size and spectrum from the matrix $A$.

\section{Conclusion}

The study of mass-spring chains as classical analogs of quantum spin systems with perfect state transfer and fractional revival has been pursued here. This has been done with the goal of adding to the two known analytical models which are connected to families of orthogonal polynomials. The first system \cite{Vaia_NewtonCradle} is associated to the dual Hahn polynomials and exhibits perfect return. This mass-spring chain was shown in \cite{SVZ} to be a special case of models based on the para-Racah polynomials and these more general chains were found to have both perfect transfer and fractional revival.

We have described in this paper novel models, with again both these peoperties, that can be designed with the help of special $q$-Racah polynomials. The use of the family of polynomials that sits at the top of the discrete part of the $q$-Askey scheme is of interest on its own and it should be noted that the appropriate specialization arose from the requirement that the recurrence coefficients be mirror-symmetric.

Mass-spring chains are basic modeling tools. The correspondance with spin chains that have been much studied in connection with quantum information tasks for instance, suggests that it would be worthwhile to keep exploring the translation in the classical realm of analyses bearing for example on almost perfect transfer \cite{Vinet_APST} or walks on graphs in the quantum domain. We plan on pursuing this.

\section*{Acknowledgements}

H.S. benefitted from an Undergraduate Student Research Awards (USRA) scholarship from the Natural Sciences and Engineering Research Council of Canada (NSERC). The research of L.V. is supported in part by a Discovery Grant from NSERC. The work of A.Z. is funded by the National Science Foundation of China (Grant No.11771015). A.Z. gratefully acknowledges the hospitality of the CRM over an extended period and the award of a Simons CRM professorship.












\end{multicols}

\end{document}